\documentclass[aps,pre,reprint,floatfix]{revtex4-1}
\usepackage{graphicx}
\usepackage{color}

\newcommand\et{ \boldsymbol{\hat{t}} }

\newcommand{\beq}{\begin{equation}}
\newcommand{\eeq}{  \end{equation}}
\newcommand{\beqa}{\begin{eqnarray}}
\newcommand{\eeqa}{  \end{eqnarray}}

\newcommand\ds{\displaystyle }

\renewcommand\et{\epsilon_{\theta}}
\newcommand\es{\epsilon_s}
\newcommand\Ecs{\mathcal{E}_s}
\newcommand\Ect{\mathcal{E}_{\theta}}
\newcommand\dint{\int \!\!\! \int}

\usepackage{xcolor}
\definecolor{red}{rgb}{1,0,0}
\definecolor{blue}{rgb}{0,0,1}

\begin{document}

\title{Impact on floating membranes}


\author{Nicolas Vandenberghe}
\email{vandenberghe@irphe.univ-mrs.fr}

\author{Laurent Duchemin}
\email{duchemin@irphe.univ-mrs.fr}

\affiliation{Aix Marseille Universit\'e, CNRS, Centrale Marseille, IRPHE UMR 7342, F-13384, Marseille, France}

\begin{abstract}
When impacted by a rigid object, a thin elastic membrane with negligible bending rigidity floating on a liquid pool deforms. Two axisymmetric waves radiating from the impact point propagate. In the first place, a longitudinal wave front -- associated with in-plane deformation of the membrane and traveling at constant speed -- separates an outward stress free domain with a stretched but flat domain. Then, in the stretched domain a dispersive transverse wave travels at a wave speed that depends on the local stretching rate. We study the dynamics of this fluid-body system and we show that the wave dynamics is similar to the capillary waves that propagate at a liquid-gas interface but with a surface tension coefficient that depends on impact speed. We emphasize the role of the stretching in the membrane in the wave dynamics but also in the development of a buckling instability that give rise to radial wrinkles. 
\end{abstract}


\maketitle


\section{Introduction}
\begin{figure*}
\includegraphics{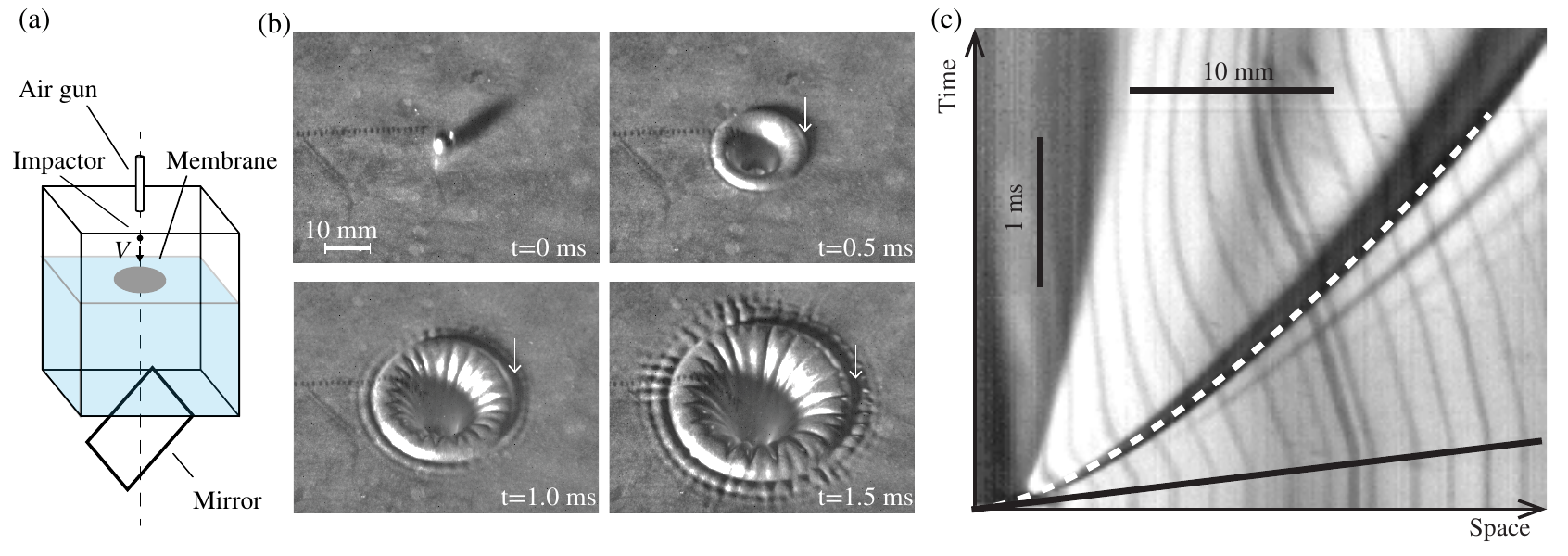} 
\caption{\label{fig:picts} 
(a) Sketch of the experiment. (b) Successive snapshots of a floating membrane (thickness $h = 0.2$ mm) made of natural rubber impacted by a solid sphere. The extension of a cavity delimited by a transverse wavefront is accompanied by the development of a wrinkling instability both inside and outside the cavity.
(c) A spatiotemporal diagram of the motion of the material points along a radius of a membrane impacted at $V=31.2$ m/s obtained by recording the movie from below the surface. Marks drawn on the membrane reveal the motion of the material points. Two waves can be seen: a longitudinal wave separates a domain where material points are at rest from a domain where they move towards the impact point.  The longitudinal wavefront travels at the speed $c = 60 $ m/s (solid line). The dark shadow highlighted by the dashed white line corresponds to the positions marked by the arrow in (b). This transverse wave front travels at non constant speed. The dashed line is the line $a' t^{2/3}$ with $a' = 13.2$ mm ms$^{-2/3}$. }
\end{figure*}

The impact of a solid sphere on a liquid surface is a scientific problem that has attracted a considerable interest since the early observations of Worthington \cite{Wor1908} of the dynamics of the cavity formed in the wake of the sphere. This problem has recently attracted a renewed interest and the cavity dynamics has been studied in detail by various authors \cite{DCD2007,AB2008}. Water entry is indeed an important problem for various applications, including aerospace science \cite{SM2006} where plane impacts have served as an early motivation \cite{K1929,W1932}, naval engineering \cite{F1993} or more recently animal locomotion \cite{BH2006}. 

In most of the applications cited above, the force exerted on the impacting body by the fluid results mainly from an inertial response. This means that other types of effects such as surface tension, gravity or viscosity are negligible. For an object of characteristic size $r_i$, impacting on a liquid surface (density $\rho$, kinematic viscosity $\nu$, surface tension $\sigma$) with a velocity $V$, the Weber number that measures the ratio between inertia and surface tension $We=\rho r_i V^2/\sigma$ and the Reynolds number $Re=V r_i/\nu$ are very large in the context of these applications. 

The case of small Weber numbers -- \textit{i.e.} surface tension dominated impacts --  has also been studied in the context of the locomotion of small organisms \cite{BH2006,VM2007}. Small Weber numbers are obtained for small objects or when the surface tension is large. The latter situation is encountered in particular when an elastic membrane is at the interface between the liquid and the gas as for a liquid filled rubber balloon \cite{LD2014}. In this case, the tension in the thin rubber membrane results from the internal pressure in the balloon and it can be several orders of magnitude larger than the typical value for the water/air interface. Provided that the initial tension is large and the amplitude of the motion of the surface of the balloon is small, the analogy with a liquid-gas interface is straightforward. However, in the general case, the tension in the membrane varies with its local strain typically through Hooke's law. Therefore the surface tension coefficient is in general not uniform: it is a dynamical variable that changes with the longitudinal (or tangential) motion of the membrane. Similar situations are encountered in the case of liquid-gas interfaces in the presence of surfactants, where longitudinal motion at the interface may change the local concentration of surfactants and thus the surface tension \cite{L1968a,L1968b,CCR1989,C2005}. 

In the present paper, we investigate a situation in which a floating membrane acts locally like a liquid-gas interface with a very large surface tension coefficient compared to a liquid-gas interface, such that, for our experiments, the Weber number is smaller than $1$. The Reynolds number is still very large, which allows us to neglect viscous effects. Gravity can also be neglected. The membrane is initially flat and tension-free (the small background stress imposed by the liquid-air surface tension at the side of the membrane can be neglected). In this case the surface tension coefficient is not known \textit{a priori}, and it results from a coupling between the transverse motion (throughout the text, transverse refers to the direction normal to the initially flat interface) resulting from the impact and the stretching of the membrane that builds up following the impact. Because of this coupling and because the background stress is negligible, the problem is considerably more difficult than the classical problem of wave propagation on stretched membranes \cite{Graff1975}. In the absence of a liquid substrate, the dynamics has been investigated by various authors in the context of ballistic impacts \cite{RD1966,PP2003,VVV2008}. These studies have shown that the displacement in the plane of the membrane and the transverse displacement are strongly coupled and need to be determined concomitantly. Furthermore the analysis has shown that the axisymmetric wave propagation can be unstable and that wrinkles resulting from a buckling instability appear. 

In a recent paper, we have investigated the two-dimensional problem \cite{DV2014} of a thin floating membrane, made of natural rubber, being impacted by a horizontal metal rod. In this context, the stretching response of the membrane is of the same nature as liquid-gas surface tension, but shows a surface tension coefficient increasing linearly with the impact velocity. This two-dimensional coupled dynamics between the membrane and the liquid (water) on which it floats, exhibits self-similar solutions like the ones observed for a liquid-gas interface \cite{KM1983}. This surprising fact results from the uniform state of strain in the membrane, allowing for a straightforward analogy between an impacted membrane and a fluid interface. In the present paper, we investigate axisymmetric situations, for which the simplification of a uniform membrane stress does not hold. Indeed, when a metal ball or a vertical cylinder impacts the membrane, the resulting strain in the membrane, and therefore the related stresses, are highly non-uniform, as we shall see.

The paper is organized as follows: in section \ref{sec:exp} we describe the experiments and the phenomenology of the impacted membrane: in particular we emphasize the propagation of two distinct waves, a wave associated with the in-plane motion of material points of the membrane, that is decoupled from the hydrodynamics, and a transverse wave that is accompanied by fluid motion. The wave dynamics is addressed in details in section \ref{sec:waves}, then the static membrane equations are solved numerically in section \ref{sec:strain} in order to understand the scaling for the strain observed in the experiments. Section \ref{sec:wrinkling} discusses the wrinkling instability of the membrane that develops as the waves propagate and section \ref{sec:deceleration} discusses the deceleration of the impactor. 

\section{Experiments \label{sec:exp}}
%

\subsection{Setup}

A steel impactor with a hemispherical head of radius $r_i = 0.75$ or $1.5$ or $2.5$ mm, impacts transversally a thin rubber membrane (thickness $H$ in the range $0.15$ to $0.30$ mm) floating at the surface of a water tank of dimensions $40 \times 40 \times 40$ cm$^3$. Impactors are accelerated by gravity or by a gas gun and they impact the membrane at speeds in the range $0.3$ to $30$ m/s. Experiments addressing the wave dynamics (in particular at low impact speeds) are conducted with impacting cylinders, that are sufficiently heavy to ensure that the impactor does not decelerate during the experiment. The membrane is a circular or square sheet of typical dimension (diameter or side length) of $15$ cm. The shape of the membrane is not relevant since we focus on short-time dynamics, before waves have had time to interact with the boundary. The typical duration of an experiment is less than $10$ ms and the dynamics is recorded with a high-speed camera at typical frame rates of 10000 frames per second. 

The membrane is characterized by a stretching modulus $Y= E' H$ with $E' = E/(1-\nu^2)$ where $E$ is Young's modulus, $\nu=1/2$ is the Poisson ratio and $H$ the thickness of the membrane in the reference state. The natural rubber used in the experiments has a Young modulus $E=2.6$ MPa, and a stretching rate of 500$\%$ or more can be reached. For simplicity and since most of the dynamics occurs at low stretching (below 10$\%$ except in a small area near the impactor), we will assume that the material response can be accurately described by Hooke's law (see \cite{AR2014} for references to more complex models in the context of impacts). A typical value of the stretching modulus is $720$ N/m. Thus, even for low  strains $\epsilon$ of the order of $10^{-3}$, the stresses induced by strains $Y \epsilon$ are well above the stress resulting from surface tension at the liquid-air interface and the latter will be neglected throughout this work. 

\subsection{Phenomenology}

After impact, two waves propagate on the membrane (figure~\ref{fig:picts}).
They are clearly seen on a spatiotemporal diagram showing the position of material points as a function of time. A material point located at a distance $R$ from the impact point is first reached by a longitudinal wave front. Behind this wavefront, the material point moves in the plane of the membrane towards the impact point. The displacement is radial and is denoted $u(R,t)$. At later times, the material points move in the transverse direction, the transverse displacement being denoted $w(R,t)$, experiencing first an oscillation of growing amplitude and later a strong transverse motion first upward and then downward towards the liquid. The wavefront associated with the longitudinal motion travels at constant speed (figure \ref{fig:picts}b). Behind this longitudinal wavefront, the membrane is stretched in a non-uniform manner: the stretching increases towards the impact point. The transverse wavefront travels in the stretched domain. This out-of-plane displacement occurs in an area well delimited by a hump, which we use to define $r_f(t)$ (see figure \ref{fig:sketch}). The transverse wavefront travels at a speed that decreases with time and its position is well approximated by the law $r_f (t) \sim a t^{2/3}$ (figure \ref{fig:picts}b). This scaling is typical of surface tension driven flows and it has been observed for two dimensional impacts on membranes \cite{KM1983,VM2007,DV2014}. The coefficient $a$ changes with the impact speed. This wave dynamics will be discussed in the next section.   

During its extension, the axisymmetric wave pattern presents an instability and radial wrinkles appear, as seen in figure~\ref{fig:picts}a. Such patterns appear frequently on elastic membranes which are not able to withstand compressive in-plane stresses \cite{CM2003,HJJ2007}. Inside the cavity, wrinkles develop on the curved membrane. The number of wrinkles does not change as the cavity extends and the wrinkles extend from the vicinity of the contact with the indenter to the ridge of the cavity. Outside of the cavity, radial wrinkles are also present. The number of wrinkles outside the cavity is different from the number of wrinkles inside the cavity. We note that there is a transition area between the cavity and the outer domain on which wrinkles are not observed. 

\section{Wave dynamics \label{sec:waves}}

\subsection{The response of a membrane to transverse impact}
\begin{figure}
\includegraphics{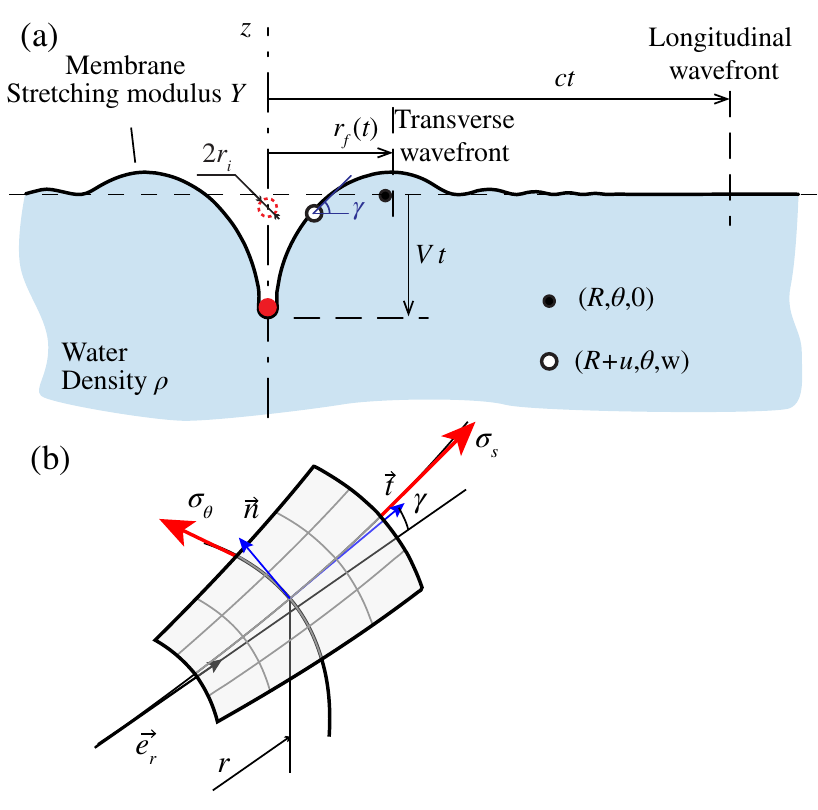} 
\caption{\label{fig:sketch} 
(a) Sketch of the impacted membrane and definition of the length $r_f$. 
(b) Notations for the model.  }
\end{figure}

We make the usual assumptions of wave theory (inviscid and irrotational flow of an incompressible fluid) for which we can use the velocity potential $\phi(r,z,t)$. We assume that the membrane plays the role of an interface and that it can be described as an infinitely thin sheet whose position is denoted $w(r,t)$ (figure \ref{fig:sketch}). 
At this stage of the problem we assume that the problem is axisymmetric. The velocity potential verifies the continuity equation $\Delta \phi = 0$ in the fluid domain. At the interface, $z=w(r,t)$, the kinematic boundary condition reads 
\begin{equation} \label{eq:kin}
\partial_t w + \partial_r w \, \partial_r \phi = \partial_z \phi.
\end{equation}
The dynamic boundary condition (gravity is neglected) reads
\begin{equation} \label{eq:ber}
\partial_t \phi + \frac{1}{2} | \nabla \phi | ^2 + \frac{p}{\rho}  = 0 ,
\end{equation}
where $p$ is the pressure difference across the membrane. 

In the reference (undeformed) configuration, the membrane is flat  and a material point has coordinates $(R,\theta,0)$ in cylindrical coordinates (black dot in figure \ref{fig:sketch}a). After deformation, the position is denoted by ($r$, $\theta$, $z$) (white dot in figure \ref{fig:sketch}a). The displacements are $u = r-R$ and $w=z$. The equation of motion for an element $h r ds d\theta$ of a membrane of mass $\rho_s h r ds d\theta$ experiencing a pressure difference $p$ across its normal and forces per unit length $N_s$ and $N_{\theta}$ in the radial and orthoradial directions, along the radial and vertical directions read
\begin{equation} \label{eq:u}
\rho_s h r \frac{\partial^2 u}{\partial t^2} = - p r \sin \gamma + \frac{\partial }{\partial s} \left( N_s r \cos \gamma \right) - N_{\theta},
\end{equation}
\begin{equation} \label{eq:w}
\rho_s h r \frac{\partial^2 w}{\partial t^2} = p r \cos \gamma + \frac{\partial }{\partial s} \left( N_s r \sin \gamma \right),
\end{equation}
where $s$ is the curvilinear coordinate along a meridian line, and the angle $\gamma$ is given by
 \begin{equation}
\cos \gamma = \frac{\partial r}{\partial s} , \quad \sin \gamma = \frac{\partial w}{\partial s}.
\end{equation}
The strains in the radial and orthoradial directions are 
\begin{equation}
\es = \frac{\partial s}{\partial R} -1 , \quad \et = \frac{r}{R} - 1.
\end{equation}
From the relation $r = R (1 + \epsilon_{\theta})$, one obtains after some algebra a compatibility equation
\begin{equation} \label{eq:comp}
\frac{r}{1+\et}  \frac{d \et}{d r} + \frac{1 + \et}{1+\es} \frac{1}{\cos \gamma} = 1
\end{equation} 

To describe the longitudinal wave that propagates ahead of the transverse perturbation, we first consider the case of in-plane displacement, with $w = 0$, for which the motion of the membrane is not coupled to the fluid (because viscous effects are neglected). Then $\gamma = 0$ and assuming a Hookean behaviour of the membrane,  
\begin{equation} \label{eq:hooke}
N_s = Y (\epsilon_s + \nu \epsilon_{\theta}) , \quad N_{\theta} = Y (\epsilon_{\theta} + \nu \epsilon_s),
\end{equation}
equation (\ref{eq:u}) reads, in the limit of small strains, {\it i.e.} for $|\et|=|u/R|\ll1$,
\begin{equation} \label{eq:longwaves}
\frac{1}{c^2} \frac{ \partial^2 u}{\partial t^2} = \frac{1}{R} \frac{\partial}{\partial R} \left( R \frac{\partial u}{\partial R} \right) - \frac{u}{R^2}
\end{equation}
with $c^2 = (E'/ \rho_s)$. This equation describes the propagation of a longitudinal (\textit{i.e.}, in-plane) perturbation at speed $c$, which is a material constant, observed in the experiment (figure \ref{fig:picts}b). 

The full set of equations coupling the membrane and fluid equations cannot be solved analytically. To gain insight into the physics of the waves we make the following simplifications:  (i) we neglect the left hand side in equation (\ref{eq:u}), \textit{i.e.}, we consider that the in-plane stresses are at equilibrium up to $r=ct$, (ii) we also neglect the left hand side in equation (\ref{eq:w}) in comparaison with the fluid inertia (an hypothesis that is valid for waves with a wavelength larger than the thickness of the membrane). We obtain the following set of equations for the membrane
\begin{equation} \label{eq:eqt2}
p + \frac{1}{r} \frac{\partial}{\partial r} \left( N_s r \sin \gamma \right) = 0
\end{equation}
\begin{equation} \label{eq:eqs2}
\frac{d}{dr} \left[ r N_s \right] - N_{\theta} = 0
\end{equation}
Equations (\ref{eq:eqt2}) and (\ref{eq:eqs2}) are the normal and tangential equilibrium equations for the membrane element respectively. Equation (\ref{eq:eqt2}) is also a dynamic boundary condition for the fluid motion, coupled with Bernoulli equation (\ref{eq:ber}) through the pressure $p$. We note that the assumption (ii) cannot be made in the absence of fluid. This case of the impact on a free membrane has been addressed before \cite{RD1966,PP2003,VVV2008}.

It is instructive to consider the case of a constant and uniform tension $N_s$. The fluid pressure is related to the shape of the membrane through
\begin{equation} \label{eq:eqt3}
p + N_s \left( \frac{d \gamma}{ds} + \frac{\sin \gamma}{r} \right) = 0
\end{equation}
which is precisely Laplace's law. Therefore the analogy with surface-tension driven flows \cite{KM1983} is straightforward. In particular, dimensional analysis, reveals that at time $t$ after impact the characteristic lengthscale associated with the propagation of the transverse wave reads
$ a t^{2/3} \sim \left( {N_s t^2 / \rho} \right)^{1/3}$.

Before further discussing the model, we present quantitative experimental results on the propagation of waves and on the strain field in the membrane. 

\subsection{Experimental observations}

\begin{figure}
\includegraphics{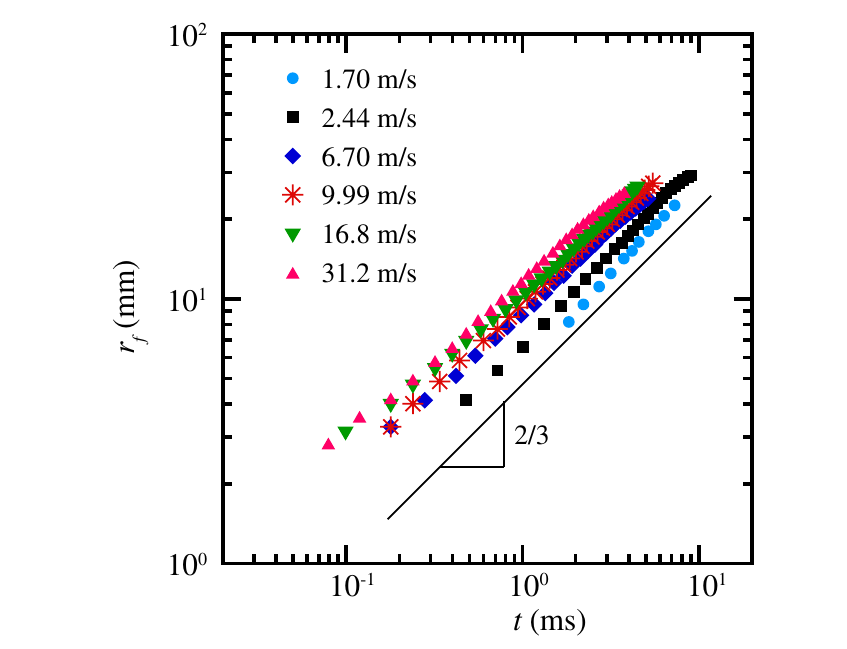}
\caption{\label{fig:Rf} 
Positions of the transverse front for different impact speeds for a membrane of thickness $h = 0.21$ mm  struck by an impactor of radius $r_i=2.5$ mm.  The scaling law $r_f = a t^{2/3}$ is robust with a coefficient $a$ increasing with the impact speed.
}
\end{figure}

\begin{figure}
\centerline{\includegraphics{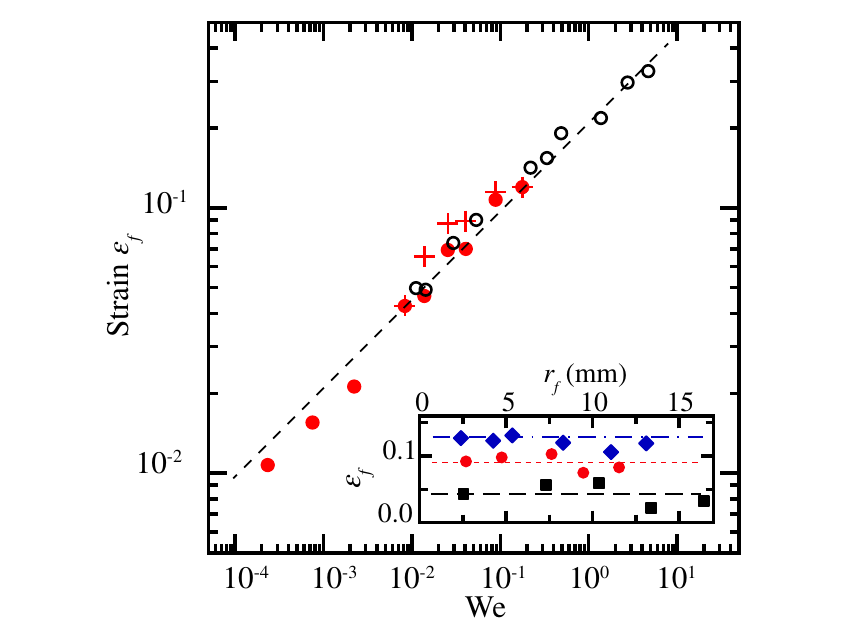}}
\caption{\label{fig:Eps} 
The strain $\epsilon_f \equiv \es (r_f)$ as a function of Weber number $We=\rho r_i V^2/Y$. The red crosses correspond to direct measurements of the strain at $r_f$ obtained by tracking the position of material points. The circles and disks correspond to the strain inferred from the prefactor $a$ obtained from the fit $r_f = a t^{2/3}$. The strain is $\epsilon_f = (1/\alpha) \rho a^3/Y$ where the coefficient $\alpha=6.1$ is chosen to match the direct measurements of $\epsilon_f$. Red disks correspond to impacts on a membrane of thickness $h=0.14$ mm with an impactor of radius $r_i = 0.75$ mm (the same conditions as the red crosses). The black circles correspond to impacts on a membrane of thickness $h = 0.21$ mm  struck by an impactor of radius $r_i=2.5$ mm (the data of figure \ref{fig:Rf}). 
The dashed line is a fit of the experimental data $\epsilon_f = \delta We^{1/3}$, 
with $\delta = 0.22$. 
The inset shows the direct measurements of the strain in $r_f$ for different times (and thus different $r_f$) at different impact speeds (black squares $V=1.9$ m/s, red disks $V = 4.3$ m/s, blue diamonds $V= 9.0$ m/s). The strain $\epsilon_f$ is constant.  }
\end{figure}

The transverse wave front travels with a well-defined law $r_f(t)=at^{2/3}$, where the constant $a$ depends on the impact velocity (figure~\ref{fig:Rf}).
Therefore, in analogy with surface-tension-driven flows, we write for $r_f$
\begin{equation}
r_f (t) = a t^{2/3} = \left( \alpha \frac{Y \epsilon_f}{ \rho} \right)^{1/3} \, t^{2/3}
\label{eq:rf}
\end{equation} 
where $\epsilon_f = \epsilon_s | _{r_f}$ is the radial strain in $r =r_f(t)$  and $\alpha$ is a number. $Y \epsilon_f$ is the local tension  in the membrane: it plays the role of the surface tension.  Measurements of the position of the transverse wavefront (figure \ref{fig:Rf}) reveal that the prefactor $a$ is, to a good approximation, constant in time, or at least exhibits a very slow variation compared to $t^{2/3}$. Therefore $\epsilon_f$ should be constant in time. Direct measurement of the strain $\epsilon_s = (\ell - \ell_0) / \ell_0$ were performed by tracking two neighboring material points drawn on the membrane and measuring the current distance $\ell$ ($\ell_0$ is the distance in the undeformed state). The strain $\epsilon_s$ measured in $r_f$ is roughly constant (inset of figure~\ref{fig:Eps}). From these observations, we conclude that the membrane behaves locally, {\it i.e.} in $r=r_f(t)$, as a liquid-gas interface, with a surface tension coefficient $Y \epsilon_f$. 

We observe in the experiments that both the coefficient $a^3$ obtained by measuring the position of the wavefront and the strain in $r=r_f$ measured by tracking material points $\epsilon_f$  scale like $We^{1/3}$ (figure \ref{fig:Eps}) where 
\begin{equation}\label{eq:We}
We = \frac{\rho r_i V^2}{Y},
\end{equation}
is the Weber number. As in the two-dimensional case \cite{DV2014}, the local strain $\epsilon_f$ depends on the impact velocity and we shall study this dependence in section \ref{sec:strain}.

The experiments also show that, for sufficient impact speeds, or for long times, the transverse displacement $w(r,t)$ is a self-similar profile of the form
\begin{equation}
w(r,t) =  \eta r_f(t) W \left( \frac{r}{r_f(t)} \right)
\label{eq:wselfsimilar}
\end{equation} 
where $\eta$ depends on impact speed $V$. This scaling is also characteristic of surface tension driven flows \cite{KM1983}.
\begin{figure}
 \includegraphics{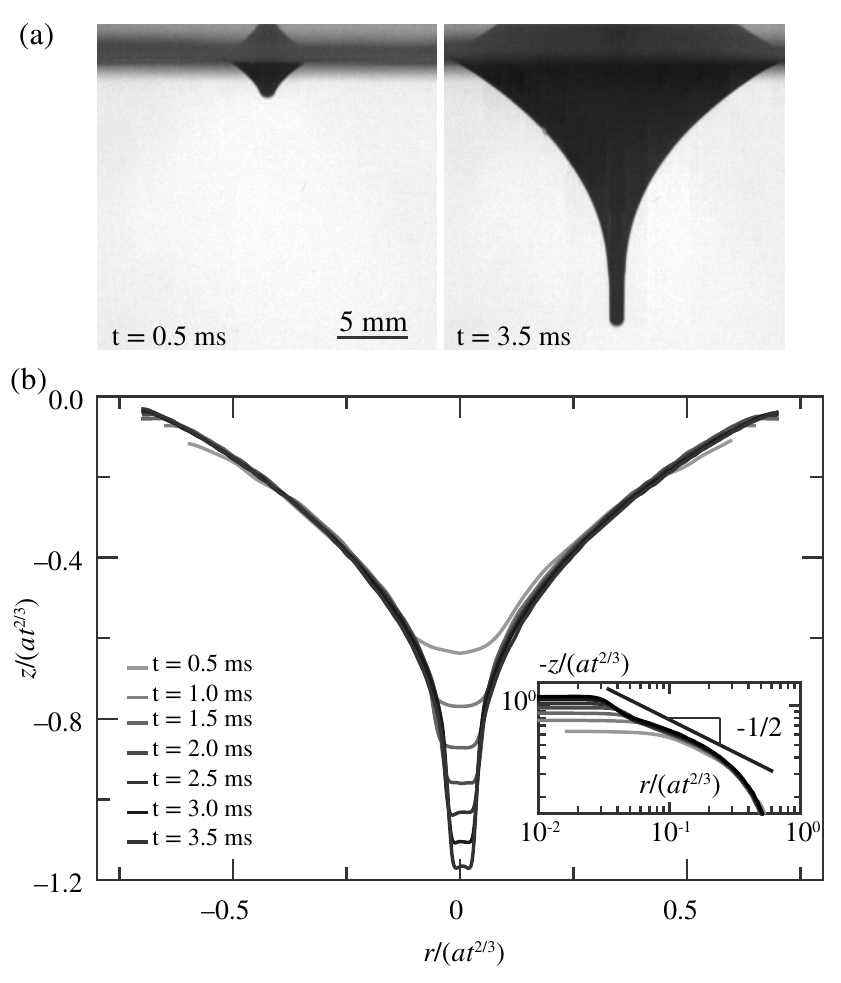}%
 \caption{\label{fig:selfsim} 
 (a) Shape (profile) of the membrane for an impactor of radius $r_i=0.5$ mm impacting at $V=5.6$ m/s a membrane of thickness $h=0.2$ mm.
 (b) Experimental profiles for which $r$ and $z$ coordinates have been rescaled by $a t^{2/3}$ with $a=7.3$ mm/ms$^{2/3}$ show a self similar behaviour. 
 Inset: The self-similar profiles shown in log-log scales. The matching of the self-similar profile with the boundary condition in $r=r_i$ imposes the shape of the profile $w \sim r^{-1/2}$ for $r \ll 1$.  }
 \end{figure}
Figure \ref{fig:selfsim} shows experimental profiles of one experiment, rescaled according to $a t^{2/3}$ (with $a$ determined on the experiment) in both directions $r$ and $z$. A unique curve is obtained away from the impactor. In the impactor region, the scaling of equation (\ref{eq:wselfsimilar}) does not hold, since the vertical displacement is $V t$, neglecting the deceleration of the impactor. 
The matching condition at the impactor $-V t = w(r_i,t)$ imposes the behaviour of the function $W$ for small $r$. To obtain linearity with time, it follows that $W(x) \sim -x^{-1/2}$ and choosing $\eta=V r_i^{1/2}/ a^{3/2}$, 
we find for $r_i/r_f \to 0$
\begin{equation} \label{eq:bcri}
w(r_i,t) \simeq  - \frac{V r_i^{1/2}}{ a^{3/2}}  \frac{r_f}{(r_i/r_f)^{1/2}} =  -Vt
\end{equation}

\begin{figure}
\includegraphics{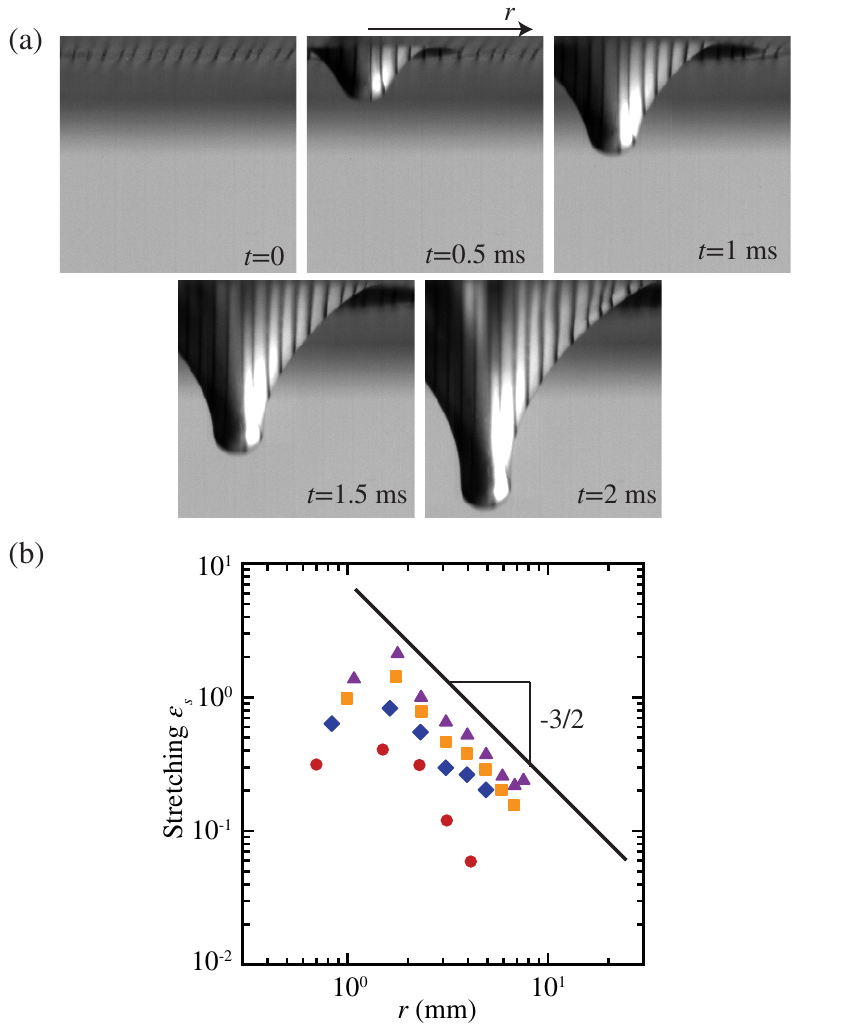} 
\caption{\label{fig:epsr} 
(a) Successive profiles showing the motion of material points. The stretching is higher near the impactor. In this experiment, the radius of the impactor is $r_i=1.5$ mm, the thickness of the elastic sheet is $h=0.2$ mm, and the impact velocity is $V=6.6$ m/s corresponding to $We = 9.1 \times 10^{-2}$.
(b) Variations of the stretching $\epsilon_s$ with $r$ for the pictures of panel (a). Time increases from bottom to top. }
\end{figure}

Despite these observations, the complete analogy with surface-tension-driven flows is not straighforward because the strain $\epsilon_s$ in the membrane is not uniform, as seen in figure \ref{fig:epsr}. Indeed, the stretching $\epsilon_s$ is large near the impactor and decreases rapidly as $r$ increases.
Thus a detailed description of the full dynamics, including the fluid and the membrane must be sought for to obtain a description of the wave dynamics and in particular to address the variation of the coefficient $a$ with impact speed. 

\section{Scaling for the strain \label{sec:strain}}

We first remark that the the full system of equations (\ref{eq:kin}, \ref{eq:ber}, \ref{eq:eqt2} -- \ref{eq:comp}), together with mass conservation in the fluid bulk $\Delta \phi = 0$, can be written using self-similar ansatz for which all the terms balance in the equations, appart from the boundary condition in $r=r_i$. The ansatz for the transverse displacement and the two strains are
\begin{equation} \label{eq:scaling1} 
w(r,t)  =  \eta r_f W (\xi), \quad \es  =  E_s (\xi), \quad \et  =  E_\theta (\xi),   
\end{equation}
where $\xi = r/r_f$ and $r_f = a t^{2/3}$. The use of the same self-similar ansatz for all the fields is dictated by the fact that the strain $\epsilon_s$ has to be a constant in the region of $r_f$ in order for $a$ to be a constant. Using the fact that $\tan \gamma = \partial w /\partial r = \eta W'(\xi)$,  the full system of self-similar equations is written in appendix \ref{app:selfsimilar}. 

This reduction to a self-similar system provides a clue as to why a scaling $r_f = a t^{2/3}$ is observed. However there are a few difficulties associated with this approach. First, the boundary conditions at $r = r_i$ and $r = c t$ are not self-similar. This means that a self-similar solution with the scalings (\ref{eq:scaling1}) will not verify the boundary conditions, and in particular the boundary condition $w(r_i,t) = -V t$ which is fundamental in the present problem. Another difficulty associated with the self-similar approach is that it does not provide a scaling for the strain. In particular, the dependence of the strain in $r=r_f$ with the impact speed cannot be determined without computing the full solution of the problem, which remains a formidable task. We propose in this section, a simplified analysis. 

In order to explore the main features of the wave dynamics -- the constant character of the strain in $r_f$ and its scaling with the impact speed -- we have solved the quasi-static membrane equations (\ref{eq:comp}) and (\ref{eq:eqs2}), using a profile $w$ compatible with the observed self-similar shape (\ref{eq:wselfsimilar}). We choose the form
\begin{equation}
w(r,t) =  - \eta r_f \frac{\sin(A r/r_f)}{A(r/r_f)^{3/2}}, 
\label{eq:memshape}
\end{equation} 
where $A=4.38$ such that the first maximum is located in $r=r_f$. Apart from the self-similarity, this shape, and especially the behavior $ w(r,t) \sim r^{-1/2}$ for small $r$, was chosen such that the boundary condition in $r_i$ (equation \ref{eq:bcri}) is satisfied. Our simplified approach consists in replacing the highly complex coupling between the membrane and the fluid motion with the feature of the flow -- built in equation (\ref{eq:memshape}) -- that the transverse wave propagates according to $r_f(t) = at^{2/3}$.
 
\begin{figure}
\includegraphics{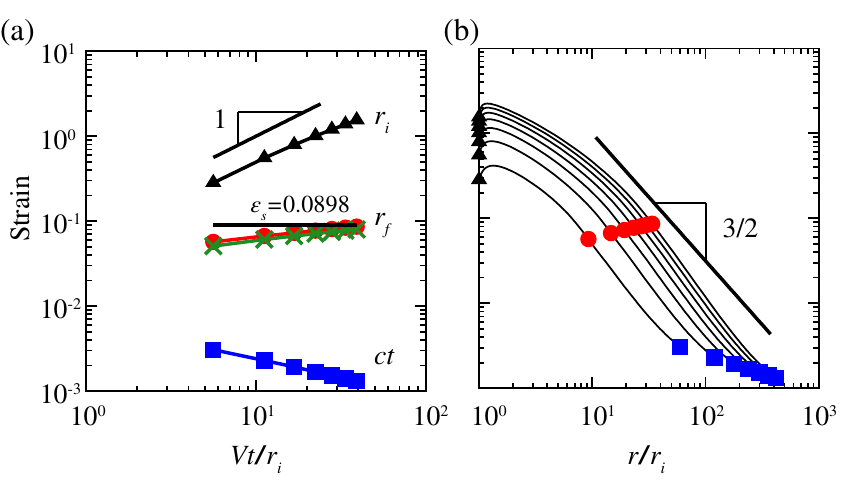}
\caption{(a) Longitudinal strains $\epsilon_s$ (triangles, discs and squares) measured in $r=r_i$, $r_f$ and $ct$, when solving equations (\ref{eq:eqs2}) and (\ref{eq:comp}), as a function of the dimensionless time $Vt/r_i$. The crosses correspond to the hoop strain $\epsilon_\theta$ in $r_f$. The parameters used in the simulation are similar to those of figure \ref{fig:selfsim}. Note that the strain $\epsilon_s$ in $r_f$ varies slowly with time and its value shows quantitative agreement with the value $9.0 \times 10^{-2}$ expected from the experiment. (b) Longitudinal strain along the membrane as a function of $r/r_i$ for times corresponding to figure \ref{fig:selfsim}.}
\label{fig:strains}
\end{figure}

\begin{figure}
 \includegraphics{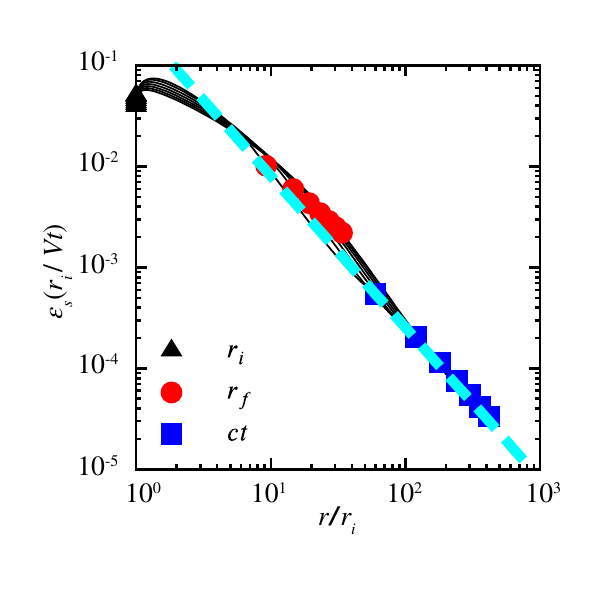}%
 \caption{\label{fig:strains_rescaled} Longitudinal strain along the membrane, rescaled according to $Vt/r_i$, as a function of $r/r_i$, when solving equations (\ref{eq:eqs2}) and (\ref{eq:comp}). The dashed line corresponds to the function $\beta (r/r_i)^{-3/2}$, where $\beta$ has been deduced from experimental measurements.}
 \end{figure}

Equations (\ref{eq:comp}) and (\ref{eq:eqs2}) were solved with a shooting technique, imposing the boundary conditions $\es(r_i)=\et(r_i)$ and $\et(ct)=0$. The solution is shown in figure \ref{fig:strains}. The same physical parameters as the experiment presented in figure \ref{fig:selfsim} have been used, and the equations were solved for the same seven times. 
Figure \ref{fig:strains}(a) shows the values of $\epsilon_s$ in $r_i$, $r_f$ and $ct$ as a function of the rescaled time $Vt/r_i$. To a good approximation, the strain $\es(r_i)$ is linear in time and $\es(r_f)$ is constant. Moreover rescaling the strain $\es$ by $Vt /r_i$ leads to the collapse of curves for different times, as seen in figure \ref{fig:strains_rescaled}. In the light of these results and seeking an ansatz with constant $\es$ in $r = r_f$, compatible with equation (\ref{eq:scaling1}), we choose the following representation for the strain field:
\begin{equation}
\es(r,t) = \beta \frac{Vt}{r_i} \left( \frac{r}{r_i} \right)^{-3/2}, 
\label{eq:eset}
\end{equation}
where $\beta$ is a numerical constant, $\epsilon_{\theta}$ admitting a similar form (with a different prefactor). The scaling in $(r/r_i)^{-3/2}$ agrees with the results in figure \ref{fig:strains}(b) and \ref{fig:strains_rescaled} and with the measurements presented in figure \ref{fig:epsr}. It is also compatible with the self-similar scaling of equation (\ref{eq:scaling1}): taking $t = (r_f / a)^{3/2}$ in equation (\ref{eq:eset}) yields
\begin{equation} \label{eq:esrf}
\es(r,t) = \beta \frac{V r_i^{1/2}}{a^{3/2}} \left( \frac{r}{r_f} \right)^{-3/2},
\end{equation}
which agrees with the form (\ref{eq:scaling1}).

Moreover, we can use equation (\ref{eq:esrf}) together with equation (\ref{eq:rf}) to obtain 
\begin{equation}
\epsilon_f = \left( \frac{\beta^2} {\alpha} \right)^{1/3} We^{1/3},
\end{equation} 
which is not only compatible with the experimental fit $\epsilon_f = \delta We^{1/3}$ as a scaling law -- as seen in figure \ref{fig:Eps} -- but also in a quantitative manner through the prefactor. This fact is clearly confirmed in figure \ref{fig:strains_rescaled}, where the prefactor $\beta$ has been determined as $\beta = \sqrt{\delta^3 \alpha}$, where $\alpha=6.1$ and $\delta=0.22$ are inferred from the experiments (see figure \ref{fig:Eps}). Equation (\ref{eq:eset}) with this value for $\beta$ fits well the numerical values of the strain as a function of $r/r_i$ in a wide region including $r_f$.

The form (\ref{eq:eset}) is valid on a large domain including $r_f$ but not near the impactor ($r \gtrsim r_i$) as seen in figure \ref{fig:strains_rescaled}.  Looking for a generalization of equation (\ref{eq:eset}), we postulate that the strains behave like 
$\es(r,t) = (Vt / r_i) \Ecs \left( r/r_i \right)$ and $\et(r,t) = (Vt / r_i) \Ect \left( r/r_i \right)$. We recall that these time and space dependancies are compatible with the self-similarity observed experimentally, provided the function $\Ecs$ and $\Ect$ scale like equation (\ref{eq:eset}) in a region enclosing $r_f$. These observations are consistent with the fact that the strain is constant in time in the vicinity of $r_f$. 

To conclude this section, we recall the main results. The transverse wave front observed in the experiments travels with a well-defined law $r_f(t)= ( \alpha Y \epsilon_f t^2 /  \rho )^{1/3}$, where $\alpha$ is a constant and the strain $\epsilon_f \equiv \epsilon_s(r_f)$ depends on the impact velocity through the scaling law $\epsilon_f =  \delta We^{1/3}$. The prefactor $\delta$ and the power $1/3$ are experimental observations, but are also consistant with a quasistatic membrane solution. This solution of the simplified problem is obtained by imposing a vertical displacement of the impactor $-Vt$ and a self-similar profile (\ref{eq:memshape}) compatible with the experiments. Using this analysis of the strain field, we now investigate the wrinkles observed in the experiments.

\section{Wrinkling of the membrane \label{sec:wrinkling}}
\begin{figure*}
 \includegraphics{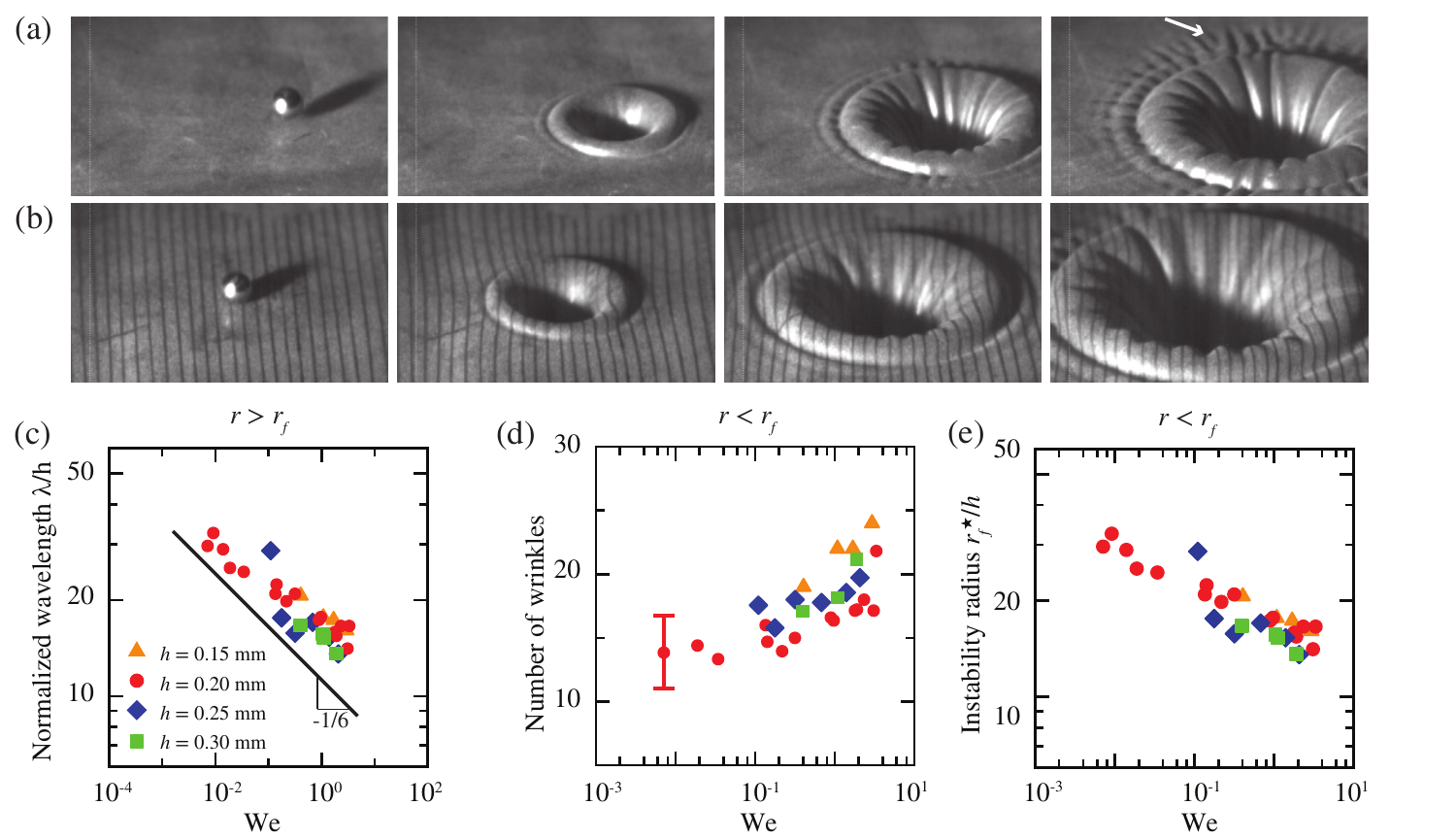}%
 \caption{\label{fig:wrinkling} 
 (a,b) Dynamics of waves and wrinkling of an impacted membrane. The experimental parameters are $r_i = 2.38$ mm and (a) $V =  26.4$ m/s, $h=0.15$ mm ; (b) $V = 30$ m/s, $h = 0.30$ mm. Pictures were taken at times: (a) $t=0$, $t=0.6$, $t=1.4$ and $t=2.2$ ms (b) $t=0$, $t=0.6$, $t=1.2$ $t=1.8$ ms. Radial wrinkles can be seen in the cavity and outside the cavity with different wavenumbers. The wrinkles outside the cavity have a well defined wavelength and as the wave propagates new wrinkles appear (see arrow). 
 (c) The wavelength of wrinkles outside the cavity agrees decreases impact speed according to equation (\ref{eq:wrinkles}).
 (d) The number of wrinkles inside the cavity is different from the number of wrinkles outside the cavity. It does not exhibit a clear variation with the sheet thickness. (e) The radius $r_f$ that has been reached by the transverse wavefront when wrinkles appear in the cavity is proportional to the sheet thickness.}
 \end{figure*}
As seen in figure \ref{fig:picts}, as the waves extend, the membrane presents an instability and wrinkles appear in two distinct domains, for $r<r_f$ where the membrane is out of its plane and is curved and for $r>r_f$ where the membrane is roughly flat (in its nominal state \textit{i.e.} before the instability) but streched. We treat these domains independently. 

\subsection{Wrinkling instability in the flat domain $r>r_f$}

The wrinkles that appear outside the cone are also observed in the absence of a liquid substrate \cite{VVV2008}. They result from a buckling instability that is caused by the motion of material points towards the impactor in the domain delimited by the longitudinal wavefront located in $ct$ and the transverse wave front located in $r_f$. As a result of the radial motion of the material points, a compressive hoop stress $\epsilon_{\theta}$ develops. This instability has been described by Vermorel et al. \cite{VVV2008}  in the absence of the liquid substrate and we adapt the analysis to describe the present case. As discussed in section \ref{sec:strain} the strain near $r_f$ is roughly constant with the scaling $\epsilon_{\theta} \sim \epsilon_s \sim We^{1/3}$ (figure \ref{fig:Eps}). We consider the simple problem of a beam of unit lateral length submitted to the hoop stress $\sigma_{\theta} = E' \epsilon_{\theta}  < 0$: the beam here represents the unfolded (\textit{i.e.} uncurved) annulus near the radius $r \gtrsim r_f$. The dispersion relation for a transverse perturbation on the beam \cite{Landau} (not accounting for added mass) is
\begin{equation}
\rho h \omega^2 = \sigma_{\theta} h k^2 + \frac{E' h^3}{12} k^4
\end{equation}
and the critical wavenumber is thus given by $h k_c \sim (|\sigma_{\theta}|/E')^{1/2}$, leading to a wavenumber of the instability
\begin{equation} \label{eq:wrinkles}
\frac{\lambda}{h} \sim We^{-1/6}
\end{equation}
This result agrees with our observations as seen in figure \ref{fig:wrinkling}c. We note also that, as in the absence of substrate, the wrinkling instability selects a wavelength, rather than a number of wrinkles. Thus as the waves propagate, new wrinkles appear (figures \ref{fig:picts} and \ref{fig:wrinkling}).

\subsection{Wrinkling instability in the cavity $r<r_f$}

Wrinkles are also observed inside the cavity unlike in the absence of a liquid substrate where they are not observed at least for moderate impact speed (compared to the speed of sound waves in the material) \cite{VVV2008, AR2014}. During an experiment the number of wrinkles $n$ observed in the cavity tends to decrease as waves propagate. The number of wrinkles is smaller inside the cavity than outside the cavity: in the last frame of figure \ref{fig:wrinkling} \textit{a}, the estimated number of wrinkles is 24 inside the cavity and 42 outside (near $r_f$). Moreover, the variations of $n$ with the Weber number and with the thickness $h$ do not agree with the scaling law (\ref{eq:wrinkles}). In particular, the variation of the wavelength with the thickness of the membrane shows a much weaker variation with the thickness $h$ if any. Such behaviour, and in particular the weak dependence of wavelength with thickness, are observed for patterns selected far above the threshold of buckling in stressed membranes \cite{GD2013}. We propose here an analysis of the pattern for a finite amplitude of the modulation of the transverse displacement. We write the transverse displacement (at a given time)
\begin{equation}
w(r,\theta) = w_0(r) + f(r) \cos (n \theta)
\end{equation} 
We consider that the transverse displacement relaxes the orthoradial strain and thus that the length of a perimeter $2\pi R$ is equal to $\int [(r d\theta)^2 + (dw)^2]$ thus yielding
\begin{equation} 
| \epsilon_{\theta} | \approx \frac{n^2 f^2(r)}{4 r^2}
\end{equation}
This is a geometrical relation between the amplitude of the pattern $f$ and its wavenumber $n$. After having experienced a buckling instability,  the membrane is bent. The change of elastic energies associated with the wrinkling take the form of a bending energy for which the strain is $(h \kappa_{\theta})$ where $\kappa_{\theta} \sim (1/r^2) \partial^2 w / \partial \theta^2$ is the curvature
\begin{equation} \label{eq:Ub}
U_b \sim  \dint Y  \left ( h^2 \frac{1}{r^2} \frac{\partial^2 w}{\partial \theta^2} \right)^2 r dr d\theta,
\end{equation}
and a stretching energy, where the excess strain in the radial direction resulting from the non-axisymmetric motion is proportional to $(\partial w / \partial r)^2$ and thus
\begin{equation}
U_s \sim \dint (Y \es) \left[ f'(r) \cos (n \theta) \right]^2 r dr d\theta.
\end{equation}
Noting $F$ the scale for the amplitude $f$ we have $U_b \sim Y h^2 F^2 n^4 / r_f^2$ and $U_s \sim Y \es F^2$ where we have assumed that the radial variations occur with a scale $r_f$. Using the relation $F^2 \sim r_f^2 | \et |/n^2$, it appears that $U_b$ scales like $n^2$ and $U_s$ like $n^{-2}$. Thus the elastic energy $U_b + U_s$ is minimal when
\begin{equation} \label{eq:n}
n^4 \sim \left( \frac{r_f}{h} \right)^{2} \es
\end{equation} 

The wavenumber that is observed in the experiment results from the pattern that develops after the onset of wrinkling. Wrinkling with a wavenumber $n$ occurs when the stretching energy associated with compression in the orthoradial direction
\begin{equation}
U_{\theta} \sim \dint (Y \et) \left[ \frac{n}{r} f(r) \sin (n \theta) \right]^2 r dr d\theta.
\end{equation}
is of the same order of magnitude as the bending energy (\ref{eq:Ub}). With the scaling $U_{\theta} \sim Y | \et |  n^2 F^2$ one obtains an instability when $r_f^2 = r_f^{\star \, 2} \sim h^2 n^2 / | \epsilon_{\theta} |$. This estimation of the characteristic time at which the pattern is selected assumes that the instability growth time is comparable with the time to reach the threshold of instability. Using $r_f^{\star}$ in the scaling (\ref{eq:n}) yields
\begin{equation}
n \sim \left( \frac{\es}{| \et |} \right)^{1/4} \quad \mathrm{and} \quad r_f^{\star} \sim h \frac{\es^{1/2}}{\et}
\end{equation}
The selected number of wrinkles observed in the experiment actually shows no clear variation with the thickness $h$, whereas the radius at which wrinkles are observed scales linearly with the thickness in agreement with the present analysis (figure \ref{fig:wrinkling}). We note however that the model does not capture the weak dependance of the number of wrinkles with the Weber number: $n$ increases with the Weber number and $r_f^{\star}$ shows a variation weaker than the expected $We^{-1/6}$. This discrepancy between the simplified model and the experimental results may be the consequence of higher order corrections in ratio $\es / \et$. The quasistatic analysis used in section \ref{sec:strain} indicates that the ratio $\es / | \et |$ increases weakly with the Weber number ($\es / | \et | \approx (1 +  s \, We^{1/3})$ where $s \lesssim 1$ is a number). 

\section{Deceleration of the impactor \label{sec:deceleration}}

\begin{figure}
 \includegraphics{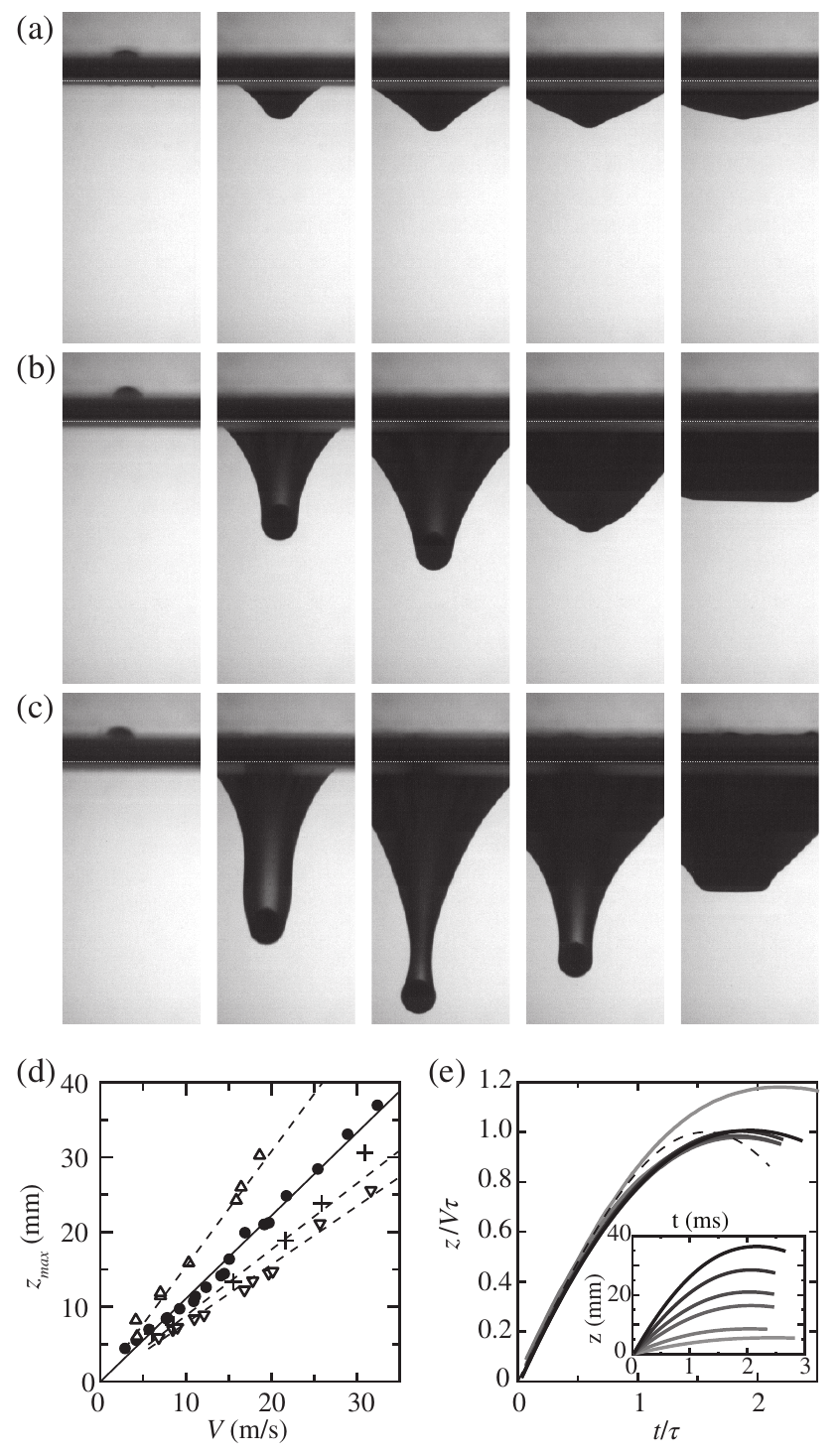}%
 \caption{\label{fig:decel} 
 Deceleration of the impactor. 
 (a-c) Images of a membrane of thickness $h=0.2$ mm as a sphere of radius $2.38$ mm and mass $m=0.44$ g impacts and decelerates. The interval between each frame is $1.07$ ms. The impact speeds are (a) $V = 6.5$ m/s, (b) $V = 19.3$ m/s, (c) $V = 32.0$ m/s.  
 (d) Maximal penetration of spheres of radius $2.25$ mm and mass $m=0.37$ g (D1, disks), $m=0.72$ g (D2, up-pointing triangles) and $m=0.18$ g (D3, down-pointing triangles) for a membrane of thickness $h=0.19$ mm and a sphere of mass $m = 0.37$ g for a membrane of thickness $h=0.29$ mm (D4, crosses). The solid line is a linear fit for the data D1, and the dashed lines are deduced from the theory (see text) for data (D2,D3,D4). 
 (e) When rescaled the raw trajectories of the impactor (shown in the inset from bottom to top data D1, with $V=4.19$, $7.80$, $15.0$, $21.0$, $25.4$, $33.0$ m/s) follow approximately the same dynamics. The dashed line shows the solution (\ref{eq:wsin}). }
 \end{figure}

The propagation of waves on the membrane and in the fluid is associated with a transfer of momentum from the impacting object. As a result, the impactor decelerates. The dynamics of the membrane as the sphere decelerates is shown in figure \ref{fig:decel}. For the three impact speeds presented in the figure, the shapes of the membrane exhibit significant differences: at low impact speed the angle $\gamma$ remains moderate, while at intermediate and high impact speeds the angle at the  contact with the impactor ($r=r_i$) reaches $\pi/2$. At high impact speeds (figure \ref{fig:decel}c) the shape of the cavity is similar to the case of a non-wetting sphere impacting a water surface (except for the absence of pinch-off) \cite{DCD2007}. This behaviour is characteristic of high Weber numbers (here $We = 3.4$). 
After impact, the sphere decelerates until it stops and then rebounds.  Figure \ref{fig:decel}d shows that the maximal penetration grows linearly with the impact speed. For moderate and high speeds, the time at which the maximal penetration occurs does not change significantly with impact speed as seen in the inset of figure \ref{fig:decel}e where the vertical position of the impactor has been plotted as a function of time, for different values of the impact speed. 

In order to make a simplified analysis of the motion of the sphere, we write the equation for the position $z$ of the impactor
\begin{equation} \label{eq:impactor}
m \frac{d^2 z}{dt^2} + F_i = 0
\end{equation}
where $m$ is the mass of the impactor and $F_i$ the force exerted by the impactor on the membrane. The simplest form for the force is to assume a quasistatic behavior of the membrane and to write the force 
\begin{equation}
F_i = 2 \pi r_i Y \epsilon_i \sin \gamma(r_i,t) \approx 2 \pi Y k z(t)
\end{equation}
where we have used the result of section \ref{sec:strain}, $\epsilon_s(r_i,t) = k V t / r_i$ with $Vt = z(t)$ and $k=\Ecs(1)$. We have also made the approximation $\sin \gamma(r_i,t) = 1$ which is valid if the impact speed is not too small. Most importantly, we have assumed that the pressure impulse occurring  just after impact could be neglected in comparaison with the tension in the membrane. This hypothesis comes from the fact that the impact force scales like $r_i^2$, whereas the tension scales like $r_i$. As a consequence, we expect $F_i$ to be much larger than the impact force, as soon as $\epsilon_i$ is non-negligeable.

This simplified model with a Hookean restoring force yields a solution, with initial conditions $z(0) = 0$ and $dz/dt(0) = V$
\begin{equation} \label{eq:wsin}
	z(t) = V\tau \sin \left( \frac{t}{\tau} \right), \quad \mathrm{with} \quad \tau = \left( \frac{m}{2 \pi Y k} \right)^{1/2}.
\end{equation}  

The time scale obtained from the experiments, as the slope of the curves in figure \ref{fig:decel}d is in fair agreement with the time scale deduced from the quasi-static model: for the data set D1, we find $\tau \approx 1.11$ ms from the experiments (solid line in figure \ref{fig:decel}d), which yields $k \approx 0.070$ in fair agreement with $k = \es(r_i,t)r_i/Vt = \Ecs(1) \approx 0.05$, as seen in figure \ref{fig:strains_rescaled}. With $k \approx 0.070$ the other data sets are well approximated by the lines $z_{max} = V \tau$ (shown as dashed lines in  figure \ref{fig:decel}d) where $\tau$ is computed from equation (\ref{eq:wsin}) with the corresponding values of the mass $m$ and thickness $h$. Solution (\ref{eq:wsin}) is in qualitative agreement with the rescaled experimental curves (figure \ref{fig:decel}e), apart from the position of the maximum. We also note that a discrepancy is observed for the lower speeds, caused by the variations of $\gamma$ not accounted for in the simplified model.  
\begin{figure}
\begin{center}
 \includegraphics{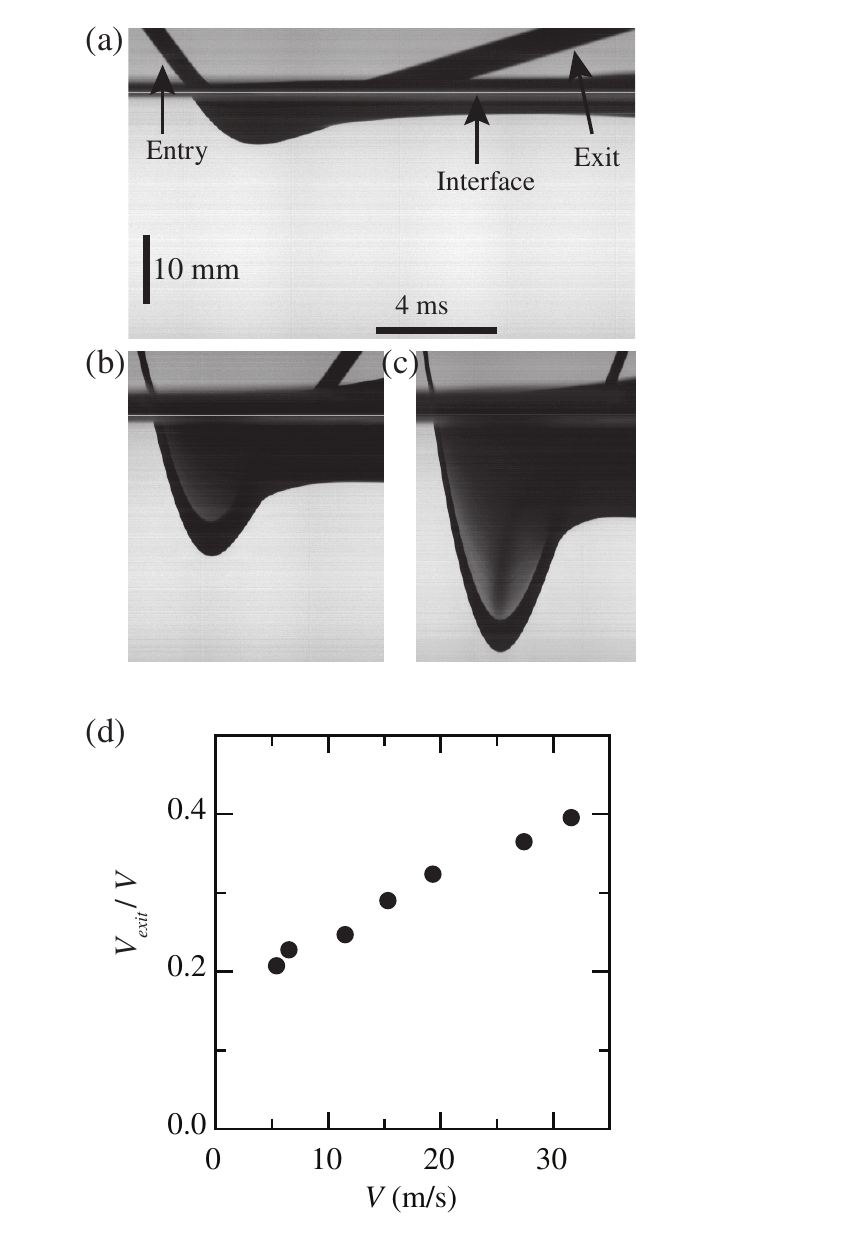}%
 \end{center}
 \caption{\label{fig:rebound} 
 (a-c) $z$-$t$ diagrams showing the entry at velocity $V$, deceleration and exits at velocity $V_{exit}$ of a sphere corresponding to the images of figure \ref{fig:decel}. 
 (d) The coefficient of restitution varies with the impact speed. }
 \end{figure}
It is also worth mentioning that after its deceleration, the sphere is accelerated by the membrane and finally ejected. Therefore we can measure a coefficient of restitution. The deceleration and rebound dynamics is presented in figure~\ref{fig:rebound}. The coefficient of restitution increases with impact speed and seems to have an asymptotic finite value as $V \to 0$. These features are also observed in the case of impact of non-wetting spheres on a water surface \cite{LK2008}. 

It is interesting to compare the dynamics of the floating membrane to the case of an impacted plate. In the present case, the stretching at the contact with the impactor, that will ultimately (for high speeds) be responsible for the puncturing of the membrane, increases progressively with time with a time scale $r_i/V$ while the deceleration occurs at a time scale $\sim (m/Y)^{1/2}$. This is different from the case of an impacted plate of thickness $h$, characterized by its bending rigidity, where the nominal curvature $V/(ch)$ is reached very rapidly, within the short time necessary to establish the Hertzian response of the plate: this time is typically $t_{H} \sim (h/c) (c/V)^{1/3} (h/r_i)^{1/3}$ \cite{Z1941}.

\section{Conclusions \label{sec:conclusion}}

We have studied the behaviour of a membrane floating on a liquid pool impacted by a rigid object. The membrane is initially stress free. Tension in the membrane develops as a result of the impact and the dynamics of the transverse wave is coupled with the tension wave. We have shown that the strain distribution -- and therefore the stress distribution -- observed in the experiments on the membrane is fully consistent with a simplified theoretical model. This model consists in the assumptions that the transverse wave front travels in the radial direction with a $t^{2/3}$ law and that the membrane is, at each instant, in internal equilibrium. This model allowed us to describe the shape of the membrane in the region of the transverse wave as a self-similar function, analog to surface-tension-dominated free-surface flows. The equivalent local surface tension coefficient, derived from the theory, is constant in time and increases like the Weber number to the power $1/3$ as observed in the experiments. Moreover, the theoretical expression for the strain in the transverse wave region gives a scaling for the wavelength of the wrinkles observed at long-time, which is in agreement with the experiments. Finally, the model allows us to understand the deceleration of the impactor~: the agreement between the theory and the experiments is fairly good in this purpose.

We leave here, as a perspective of this work, a deeper experimental study of the deceleration of the impactor and the wrinkles growth. In order to conclude on this aspect, we shall need to change the material properties (thickness of the membrane and Young's modulus, though the domain of variations is limited by the bending response that will unavoidably affects the wave dynamics for thick or rigid membranes) and the liquid properties (density, kinematic viscosity) in order to disentangle this complex long-time dynamics. One remaining open question that this future work should address is the amount of energy transfer during the impact. Indeed, quantifying the energy transferred into kinetic energy (inside the fluid) and elastic energy (in the membrane) should have many applications. 

\textbf{Acknowledgment.} We acknowledge support from the Agence Nationale de la Recherche through grant ANR-11-JS09-0005. 

\appendix

\section{Self similar system of equations}\label{app:selfsimilar}

The self-similar ansatz for the velocity potential, the pressure jump across the membrane, the vertical displacement and the two strains are 
\begin{eqnarray}
& \phi(r,z,t)  =   a^2 t^{1/3} \Phi(\xi, \zeta),\quad p(r,z,t) = \rho a^2 t^{-2/3} P(\xi,\zeta),& \nonumber\\
& w(r,t)  =  \eta a t^{2/3} W (\xi) &  \label{eq:scaling1_app} \\
&\es  =  E_s (\xi), \quad \et  =  E_\theta (\xi),&  \nonumber
\end{eqnarray}
where $\xi = r/r_f$, $\zeta = z/r_f$,  $a= (Y \epsilon_f / \rho)^{1/3}$. Plugging these expressions into mass conservation and equations (\ref{eq:kin}, \ref{eq:ber}, \ref{eq:eqt2} -- \ref{eq:comp}), we obtain the following set of equations~:
\begin{equation}
\Delta \Phi = 0
\end{equation}
for $\zeta \le W(\xi)$ and
\begin{eqnarray*}
& \ds \frac23 \eta W(\xi) - \frac23 \eta \xi W'(\xi)  =  \Phi_\zeta - \eta W'(\xi) \Phi_\xi &
\\
&\ds \frac13 \Phi - \frac23 (\xi \Phi_\xi + \eta W(\xi) \Phi_\zeta) + \frac12 \nabla \Phi^2 + P  =  0 &
\\
&\ds \frac{d}{d\xi} \left[ \xi \left( E_s + \frac{E_\theta}{2} \right) \right] - \left( E_\theta + \frac{E_s}{2} \right) = 0 &
\\
&\ds \frac{\xi}{1+E_\theta}  \frac{d E_\theta}{d \xi} + \frac{1 + E_\theta}{1 + E_s} \sqrt{1 + \eta^2 W'^2(\xi)} = 1 &
\\
&\ds \left(E_s + \frac{E_\theta}{2}\right) \frac{d \gamma}{d \xi} +\left(E_\theta + \frac{E_s}{2}\right)\frac{W'(\xi)}{\xi} \qquad&
\\
&\qquad + P \sqrt{1 + \eta^2 W'^2(\xi)} = 0 &
\end{eqnarray*}
for $\zeta =W(\xi)$
%

\end{document}